\documentclass[aps,twocolumn,showpacs,superscriptaddress]{revtex4-1}
\usepackage{graphicx}
\usepackage{rotating}
\usepackage{mathrsfs}
\usepackage{amsmath}
\usepackage{amssymb}

\begin{document}
\title{Calculation of the spectrum of $^{12}$Li by using the multistep shell model
method in the complex energy plane}
\author{Z.~X.~Xu}
\affiliation{Royal Institute of Technology (KTH), Alba Nova University Center,
SE-10691 Stockholm, Sweden}
\author{R.~J.~Liotta}
\affiliation{Royal Institute of Technology (KTH), Alba Nova University Center,
SE-10691 Stockholm, Sweden}
\author{C.~Qi}
\email{chongq@kth.se}
\affiliation{Royal Institute of Technology (KTH), Alba Nova University Center,
SE-10691 Stockholm, Sweden}
\author{T.~Roger}
\affiliation{GANIL, CEA/DSM - CNRS/IN2P3, Bd Henri Becquerel, BP 55027, F-14076 Caen Cedex 5, France}
\author{P.~Roussel-Chomaz}
\affiliation{GANIL, CEA/DSM - CNRS/IN2P3, Bd Henri Becquerel, BP 55027, F-14076 Caen Cedex 5, France}
\author{H.~Savajols}
\affiliation{GANIL, CEA/DSM - CNRS/IN2P3, Bd Henri Becquerel, BP 55027, F-14076 Caen Cedex 5, France}
\author{R.~Wyss}
\affiliation{Royal Institute of Technology (KTH), Alba Nova University Center,
SE-10691 Stockholm, Sweden}

\begin{abstract}

The unbound nucleus  $^{12}$Li is evaluated by using the multistep shell
model in the complex energy plane assuming that the spectrum is determined
by the motion of three neutrons outside the $^9$Li core. It is found that the
ground state of this system consists of an antibound $1/2^+$ state and
that only this and a $1/2^-$ and a $5/2^+$ excited states are physically
meaningful resonances.

\end{abstract}

\pacs{21.10.Tg, 21.10.Gv, 21.60.Cs, 24.10.Cn}

\maketitle

\section{Introduction}

The study of halo nuclei is one of the main subjects of research in nuclear
physics at present. Many theoretical predictions on halo, superhalo and
antihalo nuclei have been advanced in recent years~\cite{hal1,hal2,hal3}.
Most of these calculations
correspond to nuclei very far from the stability line. They are
mainly thought as a guide for experiments to be performed in coming facilities.
The general feature found in these calculations is that a necessary
condition for a nucleus to develop a halo is that the outmost nucleons move
in shells which extend far in space. That is, only a weak barrier keep the
system within the nuclear volume. These shells may be resonances, antibound
states (also called virtual states), or even low-spin bound states which lie very
close to the continuum threshold.
These conditions are fulfilled by the nucleus $^{11}$Li and also heavier Li
isotopes. There are a number of experiments which have been performed in
these very unstable isotopes in order to get information about the structure
of halos~\cite{bau07}. In particular we will concentrate our attention to Refs.~\cite{aks08,pat10,hal10} where the spectrum of $^{12}$Li was measured. Our aim is
to analyze these experimental data by using a suitable formalism to treat
unstable nuclei. This formalism is an extension of the shell model to the
complex energy plane and is therefore called complex shell model~\cite{idb02}, although the name Gamow shell model is also used~\cite{mic02}.
In addition, the correlations induced by the pairing force acting upon
particles moving in decaying single-particle states will be taken into
account by using the multistep shell model (MSM)~\cite{blo84}.

The formalism is presented in Section \ref{form}. Applications are in
Section \ref{appl} and a summary and conclusions are in Section~\ref{sumc}.

\section{The formalism}
\label{form}
The study of unstable nuclei is a very difficult undertaking since, in
principle, time dependent formalisms should be used to describe the motion
of a decaying nucleus. However, the system may be considered stationary
if it lives a long time. In this case the time dependence can be circumvented.
In fact, often unstable nuclei live a very long time and therefore they may be
considered bound as, e.g., in alpha decaying states of many heavy isotopes, like
$^{208}$Bi or $^{180}$Ta(9$^-$), with $T_{1/2}>10^{15} y$.
On the other hand, experimental
facilities allow one nowadays to measure systems living a very short time. To describe
these short time processes one has
to consider the decaying character of the system. This is shown, e.g.,
by the failure of the shell model calculation of Ref. \cite{pop85},
performed by using a standard bound representation, to
explain even the ground state of $^{12}$Li \cite{aks08}.

Of the various theories
that have been conceived to analyze unbound systems, we will apply an extension
of the
shell model to the complex energy plane~\cite{idb02}. The basic assumption
of this theory is that resonances can be described in terms of states
lying in the complex energy plane. The real parts of the corresponding
energies are the positions of the resonances
while the imaginary parts are minus twice the corresponding widths, as it was
proposed by Gamow at the beginning of quantum mechanics~\cite{gam28}.
These complex states correspond to solutions of the Schr\"odinger equation with
outgoing boundary conditions. We will not present here the formalism in
detail, since this was done many times before, e.g., in Refs.
\cite{cxsm,mic09}. Rather, we will give the main points necessary for the
presentation of the applications.

\subsection{The Berggren representation}
\label{sec:berg}

In this Subsection we will very briefly describe the representation to be
used here.

The eigenstates of a central potential obtained as outgoing
solutions of the Schr\"odinger equation can be used to
express the Dirac $\delta$-function as \cite{b68}
\begin{equation}\label{eq:delb}
\delta(r-r')=\sum_n w_n(r) w_n(r') + \int_{L^+} dE u(r,E) u(r',E),
\end{equation}
where the sum runs over all the bound and antibound states plus the complex
states (resonances) which lie between the real energy axis and the integration contour
$L^+$.  The wave function of a state $n$ in these discrete set is
$w_n(r)$ and  $u(r,E)$ is the scattering function at energy $E$.
The antibound states are virtual states with negative scattering length.
They are fundamental to describe nuclei in the Li region \cite{thz}.

Discretizing the integral of Eq. (\ref{eq:delb}) one obtains
the set of orthonormal vectors $\vert \varphi_j\rangle$
forming the Berggren representation \cite{lio96}. Since this discretization
provides an approximate value of the integral, the Berggren vectors fulfill
the relation $I\approx\sum_j \vert \varphi_j\rangle \langle \varphi_j\vert$,
where all states, that is bound, antibound, resonances  and discretized
scattering states, are included. The corresponding single-particle wave
functions are
\begin{equation}\label{eq:spwf}
\langle \vec r\vert \varphi_i\rangle
=R_{n_il_ij_i}(r)\big (\chi_{1/2}Y_{l_i}({\hat r})\big )_{j_im_i},
\end{equation}
where $\chi$ is the spin wave function and 
\begin{equation}\label{eq:rphi}
R_{n_il_ij_i}(r)
=\phi_{n_il_ij_i}(r)/r
\end{equation}
is the radial wave function fulfilling the Berggren metric, according to which the
scalar product between two functions consists of one function times the other
(for details see Ref. \cite{lio96}), i.e.,
\begin{equation}\label{eq:bergm}
\int_0^{\infty} dr \phi_{n_il_ij_i}(r)\phi_{n'_i l_ij_i}(r)
=\delta_{n_in_i'}.
\end{equation}

Using the Berggren representation one readily gets the
two-particle shell-model equations in the complex energy plane (CXSM)
\cite{cxsm}, i.e.,
\begin{equation}\label{eq:sme}
(W(\alpha_2)-\epsilon_i-\epsilon_j)X(ij;\alpha_2)=
\sum_{k\leq l}\langle\tilde k\tilde l;\alpha_2\vert V\vert ij;\alpha_2\rangle X(kl;\alpha_2),
\end{equation}
where $V$ is the residual interaction. The tilde in the interaction matrix
element denotes mirror states so that in
the corresponding radial integral there is not any complex conjugate,
as required by the Berggren metric.
The two-particle states are labeled by
$\alpha_2$ and Latin letters label single-particle
states.
$W(\alpha_2)$ is the correlated two-particle energy and $\epsilon_i$
is  single-particle energy. The two-particle
wave function is given by
\begin{equation}\label{eq:wfsq}
\vert \alpha_2\rangle=P^+(\alpha_2)\vert 0\rangle,
\end{equation}
where the two-particle creation operator is given by
\begin{equation}\label{eq:wfsq}
P^+(\alpha_2)=\sum_{i\leq j}X(ij;\alpha_2)
\frac{(c^+_ic^+_j)_{\lambda_{\alpha_2}}}{\sqrt{1+\delta_{ij}}},
\end{equation}
and $\lambda_{\alpha_2}$ is the angular momentum
of the two-particle state.

We will use a separable interaction as in Ref. \cite{ant}, which
describes well the states of $^{11}$Li.
The energies are thus obtained by solving the corresponding
dispersion relation.
The two-particle wave function amplitudes are given by
\begin{equation}
\label{eq:tpwf}
 X(ij;\alpha_2) = N_{\alpha_2} \frac{f(ij,\alpha_2)}{\omega_{\alpha_2}-
(\epsilon_i + \epsilon_j)},
\end{equation}
where $f(ij,\alpha_2)$ is the single particle matrix element of the field
defining the separable interaction and
$N_{\alpha_2}$ is the normalization constant determined by the condition
$\sum_{i \le j} X(ij;{\alpha_2})^2 = 1$.

The spectrum of $^{11}$Li was already evaluated within the CXSM including
antibound states~\cite{ant}. Here we will repeat that calculation in order to
determine the two-particle states to be used in the calculation of the
three-particle system, i.e., $^{12}$Li. For this we will use the Multistep
Shell Model Method, which we will briefly describe below.

\subsection{The Multistep Shell Model Method}
\label{sec:msm}

As its name indicates, the Multistep Shell Model Method (MSM) solves the
shell model equations in several steps. In the first step the
single-particle representation is chosen. In the second step the energies
and wave functions of the two-particle system are evaluated by using a given
two-particle interaction. The three-particle states are evaluated in terms
of a basis consisting of the tensorial product of the one- and two-particle
states previously obtained. In this and subsequent steps the interaction
does not appear explicitly in the formalism. Instead, it is the wave functions
and energies of the components of the MSM basis that replace the
interaction. The MSM basis is overcomplete and non-orthogonal.
To correct this one needs to evaluate the overlap matrix among the basis
states also. A general description of the formalism is in Ref. \cite{lio82}.
The particular system that is of our interest here, i.e., the
three-particle case, can be found in Ref. \cite{blo84}, where the MSM was applied
to study the three-neutron hole states in the nucleus $^{205}$Pb.

Using the Berggren single-particle representation described above, we will
evaluate the complex energies and wave functions of $^{12}$Li using the MSM basis states
consisting of the Berggren one-particular states, which are states in
$^{10}$Li, times the two-particle states corresponding to $^{11}$Li. Below we
refer to this formalism as CXMSM.

The three-particle energies $W(\alpha_{3})$ are given by \cite{blo84}
\begin{multline}\label{TDA3}
    (W(\alpha_{3})-\varepsilon_{i}-W(\alpha_{2}))\langle\alpha_{3}|(c^{+}_{i}P^{+}(\alpha_{2}))_{\alpha_{3}}|0\rangle\\
    =\sum_{j\beta_{2}}\left\{\sum_{k}(W(\beta_{2})-\varepsilon_{i}-\varepsilon_{k})A(i\alpha_{2},j\beta_{2};k)\right\}\\
    \times\langle\alpha_{3}|(c^{+}_{j}P^{+}(\beta_{2}))_{\alpha_{3}}|0\rangle,
\end{multline}
where
\begin{equation}
    A(i\alpha_{2},j\beta_{2};k)=\hat{\alpha}_{2}\hat{\beta}_{2}\left\{\begin{array}{ccc}
                                                                        i & k & \beta_{2} \\
                                                                        j & \alpha_{3} & \alpha_{2}
                                                                      \end{array}
    \right\}Y(kj;\alpha_{2})Y(ki;\beta_{2}),\\
\end{equation}
\begin{equation}
    Y(ij;\alpha_{2})=(1+\delta(i,j))^{1/2}X(ij;\alpha_{2}),
\end{equation}
and the rest of the notation is standard.

The matrix defined in Eq. (\ref{TDA3}) is not hermitian and the dimension 
may be larger than the corresponding shell-model dimension. This is due to the violations of the Pauli principle as well as overcounting of
states in the CXMSM basis. Therefore the direct diagonalization of Eq. (\ref{TDA3}) is not convenient.
One needs to calculate the overlap matrix in order to transform the CXMSM basis into an orthonormal set.
In this three-particle case the overlap matrix is
\begin{multline}\label{Overlap}
    \left\langle0|(c^{+}_{i}P^{+}(\alpha_{2}))^{\dag}_{\alpha_{3}}(c^{+}_{j}P^{+}(\beta_{2}))_{\alpha_{3}}|0\right\rangle\\
    =\delta_{ij}\delta_{\alpha_{2}\beta_{2}}+\sum_{k}A(i\alpha_{2},j\beta_{2};k).
\end{multline}

Using this matrix (\ref{Overlap}) one can transform the matrix determined by Eq. (\ref{TDA3})
into a hermitian matrix $T$ which has the right dimension. The diagonalization of $T$ provides the three-particle energies. The
corresponding wave function amplitudes can be readily evaluated to obtain
\begin{eqnarray}
    |\alpha_{3}\rangle&=&P^{+}(\alpha_{3})|0\rangle,\\
    P^{+}(\alpha_{3})&=&\sum_{i\alpha_{2}}X(i\alpha_{2};\alpha_{3})(c^{+}_{i}P^{+}(\alpha_{2}))_{\alpha_{3}},
\end{eqnarray}
where $P^{+}(\alpha_{3})$ is the three-particle creation operator.

It has to be pointed out that in cases where the basis is overcomplete the amplitudes
$X$ are not well defined. But this is no hinder to evaluate the physical quantities.
For details see Ref. \cite{blo84}.

The advantage of the MSM in stable nuclei is that one can study the influence of collective
vibrations upon nuclear spectra within the framework of the shell model.
Thus, in Ref. \cite{blo84} the multiple structure of particle-vibration coupled states in  odd Pb isotopes was analyzed.

But the most important feature for our purpose is that the CXMSM allows one
to choose in the basis states a limited number of excitations. This is because in the continuum
the vast majority of basis states consists of scattering functions. These do
not affect greatly physically meaningful two-particle states. That is, the
majority of the two-particle states provided by the CXSM are complex
states which form a part of the continuum background. Only a few of those
calculated states correspond to physically meaningful
resonances, i.e., resonances which can be observed. Below we call a ``resonance" only to a complex state which
is meaningful. These resonances  are mainly built
upon  single-particle states which are either bound or narrow resonances.
Yet, one cannot ignore the continuum when evaluating the
resonances. The continuum configurations in the resonance wave function are
small but many, and they affect the two-particle resonance significantly
\cite{cxsm}. That is, the important continuum configurations to induce
resonances are contained in the corresponding resonance wave functions.
Therefore, the great advantage of the CXMSM is that one can include in the basis only
two-particle resonances, while neglecting the background continuum states,
which form the vast majority of complex two-particle states.
The question is which are the  two-particle states that are indeed
resonances. This, and also the question of how to evaluate and recognize the
physically meaningful three-particle states, are addressed in the next
Section.

\section{Applications}
\label{appl}

In this Section we will apply the CXMSM formalism described above to study
the nucleus $^{12}$Li. As in Ref. \cite{ant}, we will take the core to be the
nucleus $^9$Li. This is justified since the three protons included in this
core can be considered frozen and, therefore, merely spectators~\cite{ber91}.

To evaluate the valence shells we will proceed as in Refs.
\cite{esb97,pac02,idb04} and choose as central field a Woods-Saxon potential
with different depths for even and odd orbital angular momenta $l$.
The corresponding parameters are (in parenthesis for odd $l$-values)
$a$= 0.670 fm, $r_0$ = 1.27 fm, $V_0$ = 50.0 (36.9) MeV and $V_{\rm so}$=16.5 (12.6)
MeV. As in Ref. \cite{idb04}, we thus found the single-particle bound states
$0s_{1/2}$ at -23.280 MeV  and $0p_{3/2}$ at -2.589 MeV forming the $^{9}$Li
core. The valence shells are the low lying resonances $0p_{1/2}$ at
(0.195,-0.047) MeV and $0d_{5/2}$ at (2.731, -0.545) MeV and the shell
$0d_{3/2}$ at (6.458,-5.003) MeV. This cannot be considered as a resonance,
since it is so wide that rather it is a part of the background.
Besides, the state $1s_{1/2}$ appears as an
antibound state at -0.050 MeV. We thus reproduce the experimental
single-particle energies as given in Ref. \cite{boh97}.
We also found other states  at higher energies, but they do not affect our
calculation because they are very high and also very wide to be considered as
meaningful resonances. We thus include in our Berggren representation only the antibound state
$1s_{1/2}$ and the resonances $0p_{1/2}$ and $0d_{5/2}$.

The energy of the resonance  $0p_{1/2}$ has been contested and instead the
value (0.563,-0.252) MeV was proposed \cite{aks08}. Since this is an
important quantity in the determination of the two- and three-particle
spectra, we will use both of them in our calculations.  We obtained this
$0p_{1/2}$ energy (i.e., (0.563,-0.252) MeV) by
choosing $V_0$=34.755 MeV for $l$ odd while keeping all the other parameters as before.

To define the Berggren single-particle representation we still have to
choose the integration contour ${L^+}$ (see Eq. (\ref{eq:delb})).

To include in the representation the antibound $1s_{1/2}$ state as well as the
Gamow resonances $0p_{1/2}$ and $0d_{5/2}$ we will use two different contours.
The number of points on each contour defines the energies of the scattering
functions in the Berggren representation, i.e., the number of basis states
corresponding to the continuum background. This number is not uniformity
distributed, since in segments of the contour which are close to
the antibound state or to a resonance the scattering functions increase
strongly. We therefore chose the density of points to be larger in those
segments.

\begin{figure}[htdp]
\includegraphics[scale=0.80]{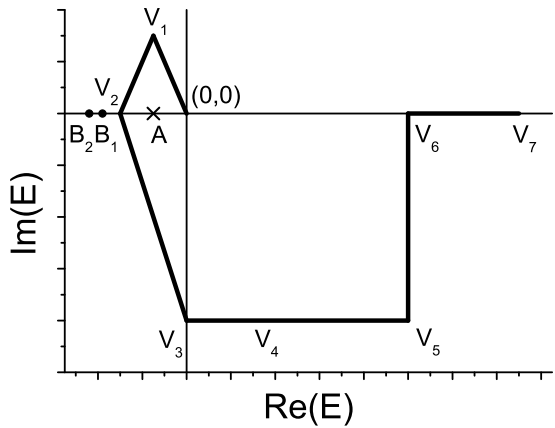}\\
\caption{Contour used to include the antibound state (see, also, Ref.~\cite{idb04}). The points
$B_i$ denote bound state energies while
$A$ denotes the antiboud state.
The points $V_i$ correspond to the vertices defining the contour.
They have the values $V_1$=(-0.05,0.05) MeV, $V_2$=(-0.1,0) MeV,
$V_3$=(0,-0.4) MeV, $V_4$=(0.5,-0.4) MeV, $V_5$=(8,-0.4) MeV, $V_6$=(8,0) MeV and
$V_7$=(10,0) MeV}
\label{cont}
\end{figure}

We include the antibound state by using the contour in Fig. \ref{cont}.

The number of points in each segment are given in Table \ref{ngp}.

\begin{table*}
  \centering
  \caption{Number of Gaussian points in the different segments of the
contour of Fig. \ref{cont}.}\label{ngp}
\begin{ruledtabular}
  \begin{tabular}{cccccccc}
Segment & [$(0,0)-V_1$]& [$V_1-V_2$]& [$V_2-V_3$]&[$V_3-V_4$]&[$V_4-V_5$]& [$V_5-V_6$]&[$V_6-V_7$]\\
Number  & 30           & 30         & 30         & 30        & 30        &    16        &   6     \\
  \end{tabular}
  \end{ruledtabular}
\end{table*}

\begin{figure}[htdp]
\includegraphics[scale=0.80]{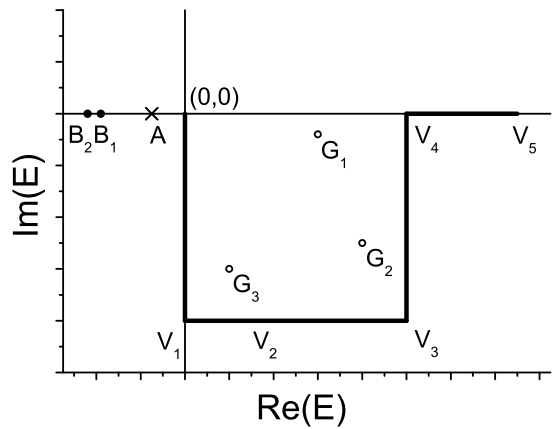}\\
\caption{Contour used to include the Gamow resonances represented by the
points $G_i$. The vertices are $V_1$=(0,-1) MeV,
$V_2$=(1,-1) MeV, $V_3$=(8,-1) MeV, $V_4$=(8,0) MeV and $V_5$=(10,0) MeV.}
\label{contr}
\end{figure}

For the Gamow resonances the contour in Fig. \ref{contr} is used with the number of Gaussian points as in Table \ref{ngr}.

\begin{table}
  \centering
  \caption{Number of Gaussian points in the different segments of the
contour of Fig. \ref{contr}.}\label{ngr}
\begin{ruledtabular}
  \begin{tabular}{cccccc}
Segment & [$(0,0)-V_1$]& [$V_1-V_2$]& [$V_2-V_3$]&[$V_3-V_4$]& [$V_4-V_5$]\\
Number  & 30           & 30         & 30         & 8        &    4      \\
  \end{tabular}
    \end{ruledtabular}
\end{table}

We have adopted these points after verifying that the results converged to
their final values.
A discussion about the choice of these contours and also on the physical
meaning of the antibound state can be found in Ref.~\cite{ant}.

With the single-particle representation thus defined we
proceed to evaluate the two-particle states.

\subsection{Two-particle states: the nucleus $^{11}$Li.}

The only state which is measured in $^{11}$Li is its bound ground state, which
was found to lie at an energy of -0.295 MeV~\cite{gar}. However,
this state is more bound than that, as it was determined in more 
recent experiments \cite{smi08,bac08,rog09}. We will adopt the most
precise of these values, i.e., -0.369 MeV \cite{smi08}. The
corresponding angular momentum is $3/2^-$. This spin arises from the odd proton
lying deep in the spectrum coupled to two neutrons. As has been already
mentioned, the proton is considered to be
a spectator. The dynamics of the system is thus determined by the pairing
force acting upon the two neutrons coupled to a state $0^+$, which behaves
as a normal even-even ground state \cite{esb97,ant}.
Besides the energy, this state has been measured to have an angular momentum
contain  of about 60 \% of  s-waves and 40 \% of p-waves,
although small components of
other angular momenta are not excluded \cite {gar}.

We will perform the calculation of the two-particle states by using the
separable interaction discussed in Section \ref{form}. The
strength $G_{\lambda_2}$, corresponding to the states with
angular momentum $\lambda_2$ and parity $(-1)^{\lambda_2}$, will be
determined by fitting the experimental energy of the lowest of these states,
as usual. It is worthwhile to point out that $G_{\lambda_2}$
defines the Hamiltonian and, therefore, is a real quantity. The two-particle
energies are found by solving the corresponding dispersion relation while the
two-particle wave function components are as in Eq. (\ref{eq:tpwf}).

\begin{table}
  \centering
  \caption{Angular momentum contain of $^{11}$Li(gs) corresponding to
  the energies $\epsilon_{p_{1/2}}$ discussed in the text. }\label{compo}
  \begin{ruledtabular}
  \begin{tabular}{cccc}
    &\multicolumn{3}{c}{component(\%)}\\\cline{2-4}
     $\epsilon_{p_{1/2}}$ (MeV) &  s-waves & p-waves & d-waves \\\hline
    (0.195,-0.047) &  46.8 & 49.1 & 4.2 \\
    (0.563,-0.252) &  72.6 & 20.9 & 6.4 \\
  \end{tabular}
  \end{ruledtabular}
\end{table}

The angular momentum contains of the ground state wave function are shown
in Table~\ref{compo}. One sees that for the case in which the
single-particle state $0p_{1/2}$ is assumed to lie at
(0.195,-0.047) MeV the two-particle wave function consists of 46.8\% $s$-states
and 49.1\% $p$-states which
are reasonable values.  For the $0p_{1/2}$ energy at (0.563,-0.252) MeV the
angular momentum contain is 72.6 \% $s$-states and 20.9 \% $p$-states,
which is also acceptable, specially considering that it provides the correct
order of the relative magnitudes. Both cases are in reasonable agreement
with experiment.

The wave function components corresponding to this state are strongly
dependent upon the contour that one uses. However, measurable quantities,
like the energies and transition probabilities, do not. This is because the
physical quantities are defined on the real energy axis and, therefore, they
remain the same when changing contour. But complex states which are part of
the continuum background do not have any counterpart on the real energy axis
and the physical  quantities for these states acquire different values for
different contours~\cite{ant}. We will use this property to determine
whether a complex state is a meaningful resonance.  This is important, since
the ground state is the only one for which experimental data exists. There
might be other meaningful states that have not been found yet. This implies
that we have to evaluate all possible two-particle states which are spanned
by our single-particle representation. 
To decide whether a state thus calculated is a meaningful resonance we will
proceed as in Refs. \cite{ant,gpr} and analyze the singlet ($S=0$) component
of the two-particle wave function.
The corresponding expression for this component  was
given in Eq. (10) of Ref. \cite{gpr}, but we will show it here again for
clarity of presentation. For the
state $\alpha$ with spin and spin-projection $(JM)$ that component is, 
with standard notation,
\begin{multline}\label{eq:wfs0}
 \Psi_{\alpha JM}(\vec{r}_1 \vec{r}_2) = 
    \left[ \chi_{1/2}(1) \chi_{1/2}(2) \right]_0^0
    \sum_{a \le b} X(ab,\alpha JM) \hat{j}_a \hat{j}_b\\
   \times \left[ C(ab,\vec{r}_1 \vec{r}_2) - (-)^{j_a+j_b-J} C(ba,\vec{r}_1 \vec{r}_2) \right]
\end{multline}
where
\begin{multline}
 C(ab,\vec{r}_1 \vec{r}_2)= \phi_a(r_1) \phi_b(r_2) (-)^{l_b+1/2-j_a+J} \\
\times  \left\{ 
   \begin{array}{ccc}
     l_a & j_a & 1/2 \\
     j_b & l_b & J \\
   \end{array}
  \right\}
  \left[ Y_{l_a}(\hat{r}_1) Y_{l_b}(\hat{r}_2) \right]_J^M,
\end{multline}
and $\phi_a(r)$ is the radial wave function corresponding to the
single-particle state $a$ (Eq. (\ref{eq:rphi})). 

If the two-particle state $(\alpha JM)$ is a meaningful resonance then the
wave function above should be localized within a region extending not too
far outside the nuclear surface, and its imaginary part should not be too large
\cite{ant}. To study these features it is not necessary to go to all six
dimensions corresponding to the coordinates $\vec r_1$ and $\vec r_2$.
In fact it is enough to consider the coordinate $r$ given by 
$\vec r_1=\vec r_2=\vec r=(0,0,r)$ which corresponds to the two particles 
located at the same point and in the z-direction. For details see
\cite{gpr}. We will call this one-dimensional function 
$\Psi_{\alpha JM}(r)$.

The evaluation of the $0^+$ states is a
relatively easy task, since in this case we have determined the strength
$G_{0^+}$ by fitting the experimental energy of $^{11}$Li(gs). With this
value of the strength we calculated all the $0^+$ states and found that there
is no any meaningful resonance. The only physically meaningful $0^+$ state is 
the bound ground state. Besides this, we found that there is a meaningful
resonance, which is the state $2^+_1$ at an energy (2.300,-0.372) MeV.
In Fig. \ref{grd} we show the corresponding radial wave function 
$\Psi_{2^+_1}(r)$. One sees that it is rather localized and that its imaginary 
part is relatively small as
compared to the corresponding real part. This is a state which perhaps is at 
the limit of what can be considered a meaningful resonance.
Yet, it is not too wide and it has an effect on the physical
three-particle states, as will be seen below. It is worthwhile to point out
that the width of this state (744 keV) is the escape width. At high
energies, where the giant resonances lie, most of the width consists of the
spreading width, i.e., of mixing with particle-hole configurations~\cite{gal88}. However, at the low energies of the states that we study this
mixing would not be relevant.

\begin{figure}[htdp]
\includegraphics[scale=0.750]{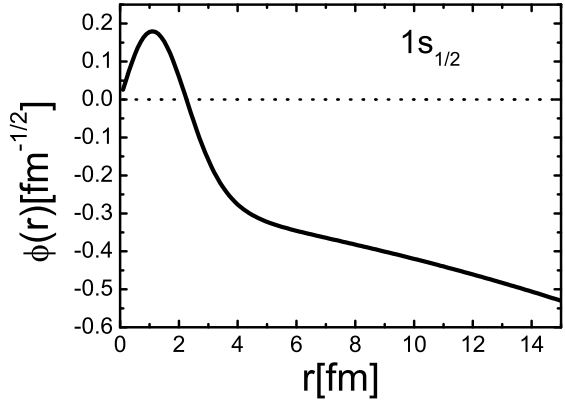}\\
\caption{Radial function $\phi(r)$ corresponding
to the single-particle neutron anti bound state $0s_{1/2}$ at an energy
of -0.050 MeV.}
\label{spwfs}
\end{figure}

\begin{figure}[htdp]
\includegraphics[scale=0.750]{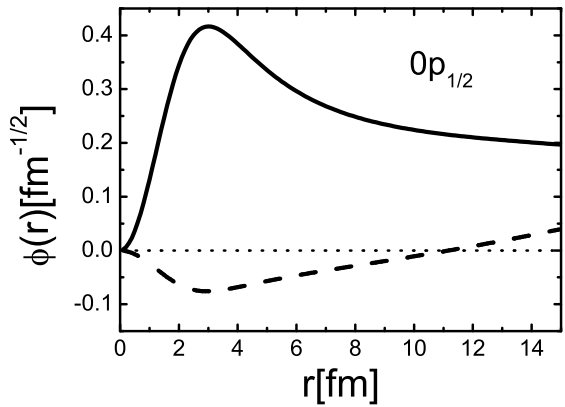}\\
\caption{As Fig. \ref{spwfs} but for the Gamow resonance
$0p_{1/2}$ at an energy of (0.563,-0.252) MeV. The dashed line
is the imaginary part of the wave function.}
\label{grp}
\end{figure}

We have thus found that the only two-particle states on which physically
meaningful three-particle states can be built are $^{11}$Li(gs) and 
$^{11}$Li($2^+_1$).
One can understand why there are so few  meaningful two-particle states in
this nucleus by looking at the  radial wave functions of the single-particle
states that form the representation. For the antibound state one sees in
Fig. \ref{spwfs} that it extends in an increasing rate far out from the
nuclear surface, as expected in this halo nucleus (the standard
value of the radius is here $1.2\times 11^{1/3}$=2.7 fm).  The radial wave 
function corresponding to the Gamow resonance $0p_{1/2}$ is shown in Fig. 
\ref{grp}.
The state $^{11}$Li($2^+_1$) is determined by the antibound state and the
$0d_{5/2}$ resonance, which has a large and increasing imaginary part
at relative short distances, as shown in Fig. \ref{grd}. These
single-particle wave functions have large imaginary parts and a divergent
nature even at rather short distances from the nucleus. In other words, they
correspond to states that live a time which is too short to produce meaningful
two-particle resonances. 

An important point for the analysis of the three-particle states to be
performed below, is that the scattering wave functions in the segments
$[(0,0)-V_1]$, $[V_1-V_2]$ and $[V_2-V_3]$ are similar in magnitude to the
wave function of the antibound state. This is because the segments are very
close to the antibound state \cite{ant}. This is a feature that cannot be
avoided, and is due to the attractive character of the pairing force. That
is, the lowest single-particle configuration in the Berggren basis is
$V_2^2$, with energy $-2\epsilon$, where $V_2=(-\epsilon,0)$. This
configuration has to lie {\it above} the energy of the two-particle
correlated state, i. e. it has to be $\epsilon>\omega(^{11}$Li(gs))/2.

\begin{figure}[htdp]
\includegraphics[scale=0.750]{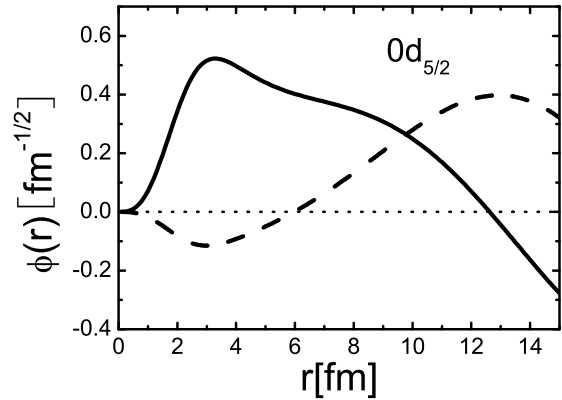}\\
\caption{As Fig. \ref{spwfs} but for the Gamow resonance
$0d_{5/2}$ at an energy of (2.731,-0.545) MeV. The dashed line
is the imaginary part of the wave function.}
\label{grd}
\end{figure}

\begin{figure}[htdp]
\includegraphics[scale=0.750]{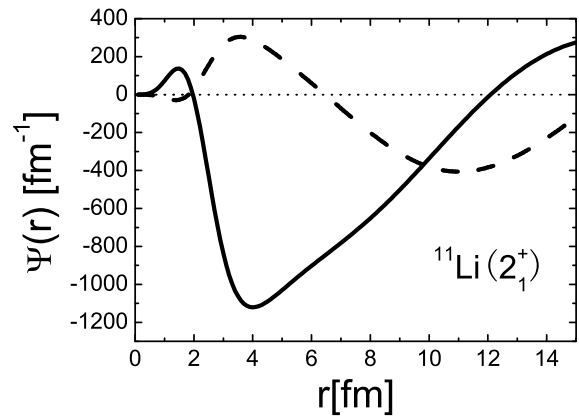}\\
\caption{Radial function $\Psi(r)$ corresponding
to the two-particle state $^{11}$Li($2^+_1$) at an energy
of (2.300,-0.372)MeV. The dashed line
is the imaginary part of the wave function.}
\label{grd}
\end{figure}

With the states $0^+_1$ and $2^+_1$ thus calculated we proceeded to the calculation of the three-particle system within the CXMSM.

\subsection{Three-particle states: the nucleus $^{12}$Li.}

With the single-particle states and the two-particle states $^{11}$Li(gs) and 
$^{11}$Li($2^+_1$) discussed above, we formed all the possible
three-particle basis states. We found that the only physically relevant
states are those which are mainly determined by the bound state $^{11}$Li($0^+_1$).
The corresponding spins and parities are 
$1/2^+$, $1/2^-$ and $5/2^+$. States like $3/2^+$, which arises from the
CXMSM configuration $|1s_{1/2}\otimes 2^+_1;3/2^+>$, is not a meaningful
state.

Due to the large number of scattering states included in the single-particle representation the dimension of
three-particle basis is also large. The scattering states are needed in
order to describe these unstable states.

In the calculations we took into account all the possibilities described
above regarding the energies of the single-particle state $0p_{1/2}$ as well
as the binding energy of the state $^{11}$Li(gs). The corresponding
results are shown in Tables \ref{Li12} and \ref{Li12s}.
Comparing these two Tables one sees that the state $1/2^+$ depends very
slightly on the energy $\epsilon_{p_{1/2}}$ but rather strongly on
${\omega_{\alpha_2}}$. Instead. this tendency is opposite for the states
$1/2^-$ and $5/2^+$.

\begin{table}
  \centering
  \caption{Calculated three-particle states in  $^{12}$Li (in MeV) corresponding
to the two energies $\epsilon_{p_{1/2}}$ of Table \ref{compo} and the 
two-particle energies taken to be ${\omega_{0^+_1}}=-0.295$MeV and
${\omega_{2^+_1}}=(2.300,-0.372)$MeV.}\label{Li12}
\begin{ruledtabular}
  \begin{tabular}{cccc}
    $\epsilon_{p_{1/2}}$ &  $1/2^+$ & $1/2^-$ & $5/2^+$ \\\hline
  (0.195,-0.047)  & (-0.386,-0.006) & (0.821,-0.189) & (1.348,-0.276) \\
  (0.563,-0.252)  & (-0.382,-0.006) & (1.114,-0.403) & (1.169,-0.242) \\
  \end{tabular}
  \end{ruledtabular}
\end{table}

The most important feature in these Tables is that the lowest state is
$1/2^+$ and that it has a real and  negative energy. It is an antibound
state, as it is the $1s_{1/2}$ state itself. A manifestation of this is that the radial
wave function diverges at large distances.

Accepting the latest reported values for the energies of the states $0p_{1/2}$
and $^{11}$Li(gs), i.e., those in the second line of Table \ref{Li12s},
our calculation predicts, besides the antibound $1/2^+$ state,
a state $1/2^-$ at (1.116,-0.411) MeV and a state $5/2^+$ at (1.148,-0.243)
MeV. This assignment agrees well with what is given in Ref. \cite{aks08}
for the ground state of $^{12}$Li, which was found to be an antibound (or
virtual) state. In the same fashion, in Ref. \cite{pat10} a state was found
at 1.5 MeV which probably has spin and parity $5/2^+$.  In these experiments the angular momentum contain of the
states were measured and therefore it is proper to compare the experimental
quantities with our calculations, where only neutrons are considered.

\begin{table}
  \centering
  \caption{As Table \ref{Li12} but for ${\omega_{0^+}}=-0.369$MeV.} \label{Li12s}
 \begin{ruledtabular}
  \begin{tabular}{cccc}
   $\epsilon_{p_{1/2}}$ & $1/2^+$ & $1/2^-$ & $5/2^+$ \\\hline
   (0.195,-0.047) &  (-0.466,-0.011) & (0.753,-0.206) & (1.315,-0.276) \\
    (0.563,-0.252)&  (-0.466,-0.011) & (1.116,-0.411) & (1.148,-0.243) \\
  \end{tabular}
   \end{ruledtabular}
\end{table}

\begin{figure*}[htdp]
\includegraphics[scale=0.550]{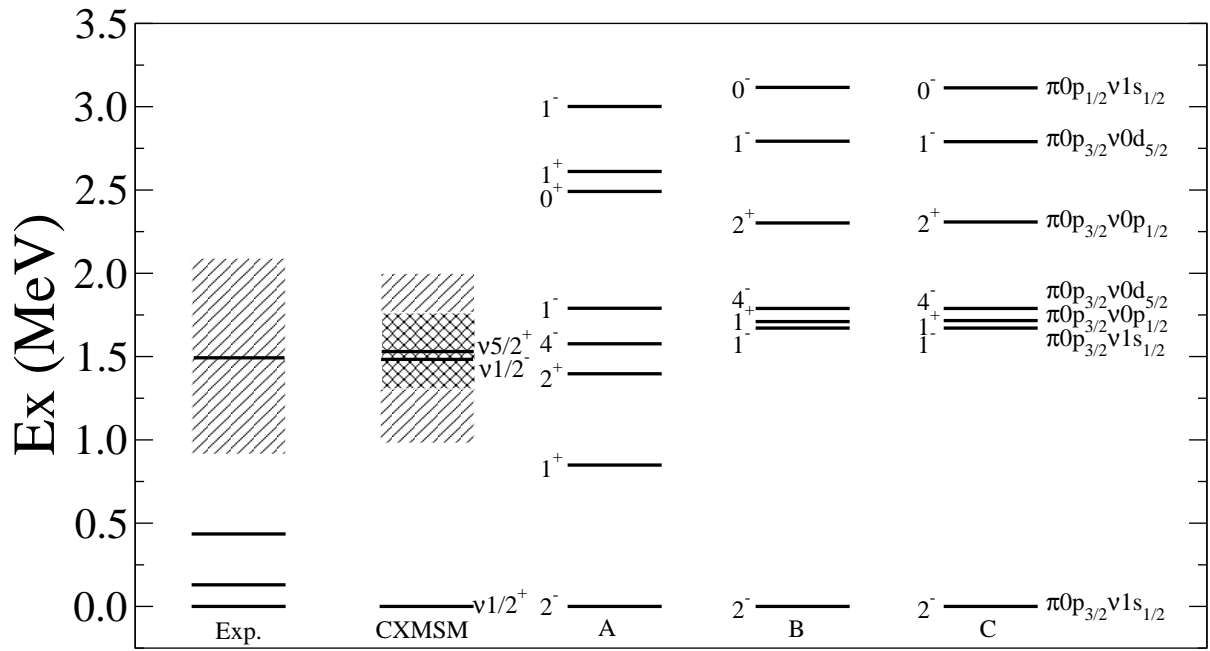}\\
\caption{ Experimental level scheme in $^{12}$Li. The three lowest levels are 
from \cite{hal10}, while the one at 1.5 MeV is from \cite{pat10}. In the second
column are the three-neutron CXMSM results.
In the columns A-C are the shell-model calculations corresponding to different truncation
schemes: A) within $0-1\hbar\omega$ excitations, B) within $0-3\hbar\omega$ excitations and
C) full $psd$ space. Dashed lines indicate the widths of the resonances.}
\label{smen}
\end{figure*}

In Ref. \cite{hal10} it was also found that $^{12}$Li(gs) is an antibound state but, in addition, two other low-lying
states were observed at 0.250 MeV and 0.555 MeV by using
two-proton removal reactions. In this case the $0p_{3/2}$ proton in
the core may interfere with the neutron excitations evaluated above.
In particular, the antibound $1/2^+$ ground state would provide, through the proton
excitation, a state $1^-$ and a $2^-$. This is the situation
encountered in the shell model calculation presented in Ref.
\cite{hal10}. As we have pointed out above, from the CXMSM viewpoint the
very unstable states determining the spectrum in this nucleus, with wave
functions which are both diverging and complex, can hardly be described by
harmonic oscillator representations. To investigate the
relation between the results provided by both representations we repeated
the shell-model calculation of Ref. \cite{hal10}. We thus took $^4$He
as the core and used the WBT effective interaction \cite{war92,bro01}. The
resulting Hamiltonian matrix was diagonalized by using the code described in
\cite{qi08}. The corresponding calculated energies, which agree with those
presented in \cite{hal10}, are shown in Fig. \ref{smen}.

One sees in this Figure that the full calculation predicts all excited states
to lie well above the corresponding experimental values. It is worthwhile to
point out that the calculated states exhibit rather pure shell model
configurations. For instance the states $2^-_1$ (ground state) and $1^-_1$
are mainly composed of the configuration
$|\pi \left[ 0p_{3/2}\right]\nu \left[ (0p_{1/2})^2 1s_{1/2}\right]\rangle$.
This does not fully agree with our CXMSM calculation, since in our case this
wave function is mainly of the form 
$|1s_{1/2}\otimes ^{11}{\rm Li(gs)}\rangle$. This differs
from the shell model case in two ways. First, the state $^{11}$Li(gs)
contains nearly as much of $1s_{1/2}$ as of $0p_{1/2}$. Second the continuum
states contribute much in the building up of the antibound $^{12}$Li(gs)
wave function, as discussed above. In our representation it is
straightforward to discern the antibound character of this state, which is
not the case when using harmonic oscillator bases. 

The shell model splittings between states originated from the same
configurations in Fig.~\ref{smen} are due to the neutron-proton
interaction, which in some cases can be large. For instance, the matrix 
elements 
$\langle\pi[0p_{3/2}]\nu[0p_{1/2}];J|V|\pi[0p_{3/2}]\nu[0p_{1/2}];J\rangle$ are
strong and attractive~\cite{tal60}.

Given all the uncertainties related to these calculations in the continuum,
and the scarce amount of experimental data,
we will not attempt to evaluate the $^{12}$Li states by adding a new
uncertainty which would be the inclusion of a proton-neutron interaction.

Considering the limitations that one expects from a calculation of
very unstable states by using harmonic oscillator basis, the rather good
agreement between the shell-model and the CXMSM presented above is
surprising. In contrast to the shell-model, the CXMSM provides not only the
energies but also the widths of the resonances. It is therefore very
important to encourage experimental groups to try to obtain these quantities
in order to  probe the formalisms.

\section{Summary and conclusions}
\label{sumc}

In this paper we have studied excitations occurring in the continuum part of
the nuclear spectrum which are at the limit of what can be observed within
present experimental facilities. These states are very unstable but yet live
a time long enough to be amenable to be treated within stationary
formalisms. We have thus adopted the CXSM (shell model in the complex energy
plane \cite{cxsm}) for this purpose. In addition we performed the shell
model calculation by using the multistep shell model. In this method of
solving the shell model equations one proceeds in several steps. In each
step one constructs building blocks to be used in future steps \cite{lio82}.
We applied this formalism to analyze $^{12}$Li as determined by the neutron
degrees of freedom, i.e., the three protons in the core were considered to
be frozen. In this case the excitations correspond to the motion of three
particles, partitioned as the one-particle times the two-particle systems.
This formalism was applied before, e.g., to study multiplets in the lead
region \cite{blo84}.

By using single-particle energies (i.e., states in $^{10}$Li) as provided by 
experimental data when available or as provided by our calculation, we found 
that the only physically meaningful two-particle states are $^{11}$Li(gs), 
which is a bound state, and $^{11}$Li($2^+_1$), which is a resonance. 
As a result there are only three physically meaningful states
in $^{12}$Li which, besides the antibound ground state, it is predicted that
there is a resonance  $1/2^-$ lying at about 1 MeV and about 800 keV wide
and another resonance which is  $5/2^+$ lying at about 1.1 MeV and 500 keV 
wide. That the ground state is an antibound (or virtual) state was confirmed 
by a number of experiments \cite{aks08,pat10,hal10} and the state $5/2^+$ has
probably been observed in \cite{pat10}. However, in \cite{hal10} two 
additional states, lying at rather low energies, have been observed which do 
not seem to correspond to the calculated levels. It has to be mentioned that 
neither a shell model calculation, performed within an harmonic oscillator 
basis, provides satisfactory results in this case. Yet, we found that this 
shell model calculation works better than one would assume given the 
unstable character of the states involved.
We therefore conclude that in order to probe the formalisms that have been introduced to describe these very
unstable systems additional experimental data, specially regarding the widths
of the resonances, would be required.

\section*{Acknowledgments}

This work has been supported by the Swedish Research Council (VR). Z.X. is 
supported in part by the China Scholarship Council under grant No. 
2008601032.

\end{document}